\documentclass[aps, twocolumn, showpacs, letterpaper]{revtex4}

\pdfoutput=1

\usepackage{amsmath}
\usepackage{amssymb}
\usepackage{graphicx}
\usepackage{xspace}
\usepackage{upgreek}
\usepackage{accents}

\newcommand{\eg}{{e.g.,\/}\xspace}
\newcommand{\ie}{{i.e.,\/}\xspace}

\newcommand{\eq}[1]{(\ref{#1})}
\newcommand{\Eq}[1]{Eq.~(\ref{#1})}
\newcommand{\Eqs}[1]{Eqs.~(\ref{#1})} 
 
\newcommand{\Ref}[1]{Ref.~\cite{#1}}
\newcommand{\Refs}[1]{Refs.~\cite{#1}}
\newcommand{\Sec}[1]{Sec.~\ref{#1}}
\newcommand{\App}[1]{Appendix~\ref{#1}}

\newcommand{\mc}[1]{\mathcal{#1}}
\newcommand{\mcc}[1]{\mathfrak{#1}}
\newcommand{\msf}[1]{\mathsf{#1}}

\newcommand{\favr}[1]{\langle #1 \rangle}
\renewcommand{\vec}[1]{{\boldsymbol{\rm #1}}}
\newcommand{\oper}[1]{\hat{\vec{#1}}}
\newcommand{\pd}{\partial}
\newcommand{\avp}{{f_s}}

\begin{document}

\title{Adiabatic nonlinear waves with trapped particles: I. General formalism}
\author{I.~Y. Dodin and N.~J. Fisch}
\affiliation{Department of Astrophysical Sciences, Princeton University, Princeton, New Jersey 08544, USA}
\date{\today}



\begin{abstract}
A Lagrangian formalism is developed for a general nondissipative quasiperiodic nonlinear wave with trapped particles in collisionless plasma. The adiabatic time-averaged Lagrangian density $\mcc{L}$ is expressed in terms of the single-particle oscillation-center Hamiltonians; once those are found, the complete set of geometrical-optics equations is derived without referring to the Maxwell-Vlasov system. The number of trapped particles is assumed fixed; in particular, those may reside close to the bottom of the wave trapping potential, so they never become untrapped. Then their contributions to the wave momentum and the energy flux depend mainly on the trapped-particle density, as an independent parameter, and the phase velocity rather than on the wave amplitude $a$ explicitly; hence, $\mcc{L}$ acquires $a$-independent terms. Also, the wave action is generally not conserved, because it can be exchanged with resonant oscillations of the trapped-particle density. The corresponding modification of the wave envelope equation is found explicitly, and the new action flow velocity is derived. Applications of these results are left to the other two papers of the series, where specific problems are addressed pertaining to properties and dynamics of waves with trapped particles.
\end{abstract}

\maketitle

\section{Introduction}
\label{sec:intro}

A standard approach to describing a nondissipative wave in the geometrical-optics (GO) limit is to start with its time-averaged Lagrangian, as proposed originally by Whitham in \Ref{ref:whitham65}; see also \Refs{book:whitham, ref:lighthill65b, ref:bretherton69, ref:dewar70, ref:hayes73, ref:kravtsov74, ref:dewar77, ref:hirota10}. Through that, both the nonlinear dispersion relation (NDR) and the action conservation theorem (ACT) are yielded, the latter being a particularly robust way to derive the envelope equation \cite{foot:robust}. However, the existing models using the time-averaged Lagrangian \cite{ref:dewar70, ref:dewar72, ref:dewar72c, ref:kravtsov74, ref:brizard95} cannot account for effects caused by particles trapped in wave troughs. In particular, those effects require special treatment for they are not necessarily perturbative, \ie may grow as the amplitude decreases \cite{my:bgk}. Hence, describing waves such as Bernstein-Green-Kruskal (BGK) modes \cite{ref:bernstein57, ref:ng06, ref:schamel00, my:bgknum} has been limited to more complicated kinetic models \cite{ref:schamel00, ref:krasovskii89, ref:krasovskii95, ref:benisti07, ref:matveev09, ref:bohm49}, which are specific to particular settings and may not render the underlying physics transparent. Therefore, it would be beneficial to generalize Lagrangian theories to accommodate trapped-particle effects.

It is the purpose of this paper to do so. Specifically, a Lagrangian formalism is developed here for general nondissipative quasiperiodic nonlinear waves in collisionless plasma, under the assumption that the number of trapped particles remains fixed. This assumption is obviously satisfied for any stationary homogeneous wave. For nonstationary or inhomogeneous waves, the number of trapped particles being fixed implies that (i) those are trapped deeply, such that they do not become untrapped when the wave parameters evolve; (ii) also, it is implied that there are no passing particles close to the resonance, so that no additional trapping can result from the wave evolution. (As models, corresponding distributions already proved useful for understanding paradigmatic effects driven by trapped particles \cite{ref:kruer69, ref:goldman71, ref:krasovsky94, ref:krasovsky09}; yet, they can also form naturally as waves evolve \cite{ref:krasovskii95}.) In particular, in the case of nonstationary or inhomogeneous waves, distributions smooth across the resonance are not~allowed. 

Under the aforementioned assumptions, we express the adiabatic time-averaged Lagrangian density $\mcc{L}$ in terms of the single-particle oscillation-center (OC) Hamiltonians; once those are found, the complete set of GO equations is derived without referring to the Maxwell-Vlasov system. Since the number of trapped particles is fixed within our model, their contributions to the wave momentum and the energy flux depend mainly on the trapped-particle density, as an independent parameter, and the phase velocity, rather than on the wave amplitude $a$ explicitly; hence, $\mcc{L}$ acquires $a$-independent terms. (Of course, taking the limit $a \to 0$ in $\mcc{L}$ would require that the width of the trapped-particle distribution also be zero, \ie that the distribution be $\delta$-shaped.) Also, the wave action is generally not conserved, because it can be exchanged with resonant waves of the trapped-particle density. The corresponding modification of the wave envelope equation, or the ACT, is found explicitly for one-dimensional (1D) waves, a case in which the trapped-particle density is expressed directly in terms of the wave variables, thus providing an exact closure. 

The results presented here extend our \Ref{my:bgk} in that we now (i)~allow plasma parameters to vary slowly in space and time, and (ii)~derive the corresponding envelope equation, or the ACT, in addition to the NDR. Applications of these results are left to \Refs{tex:myactii, tex:myactiii} (further referred to as Paper~II and Paper~III), where specific problems are addressed pertaining to properties and dynamics of waves with trapped particles. 

The paper is organized as follows. In \Sec{sec:wavelagr}, we derive the general form of $\mcc{L}$. In \Sec{sec:1dtrap}, we consider 1D waves in particular and obtain the corresponding ACT and NDR. In \Sec{sec:lwaves}, longitudinal electrostatic waves are studied as a special case. For an arbitrarily nonlinear wave, the particle OC Hamiltonian is derived, generalizing the dipole ponderomotive Hamiltonian. Then, the action density, the action flux density, and the action flow velocity are inferred. In \Sec{sec:discuss}, we also calculate those quantities specifically in the small-amplitude limit. In \Sec{sec:concl}, we summarize our main results. Some auxiliary calculations are also presented in appendixes.

\section{Wave Lagrangian}
\label{sec:wavelagr}

In \Ref{my:bgk}, we proposed the following expression for the Lagrangian spatial density of an adiabatic wave in collisionless plasma \cite{foot:similar}:
\begin{gather}\label{eq:lorig}
\mcc{L} = \favr{\mcc{L}_{\rm em}} - \sum_s n_s \favr{\mc{H}_s}_{\avp}.
\end{gather}
Here $\favr{\mcc{L}_{\rm em}}$ is the time-averaged Lagrangian density of the electromagnetic field, summation is taken over distinct species $s$, $n_s$ are the corresponding average densities, and $\favr{\mc{H}_s}_{\avp}$ are the corresponding oscillation-center (OC) Hamiltonians averaged over the distributions $f_s$ of canonical momenta $\vec{p}$. The formula was originally derived for homogeneous stationary waves \cite{foot:modes}, the case in which $\vec{p}$ and $n_s$ are constants. What we show below is that \Eq{eq:lorig} holds also in the general case, except now one needs to specify how $n_s$ relate to the field variables.

\subsection{Plasma Lagrangian}

Consider the Lagrangian $L_\Sigma = \int \mcc{L}_\Sigma\,d\msf{V}$, with the spatial density $\mcc{L}_\Sigma = \mcc{L}_{\rm em}+ \mcc{L}_p$ \cite{book:goldstein}. Here $\mcc{L}_{\rm em} = (E^2 - B^2)/(8\pi)$ is the field Lagrangian density, $\vec{E} = - \nabla \varphi - \pd_t \vec{A}/c$ is the electric field, $\vec{B} = \nabla \times \vec{A}$ is the magnetic field, $\varphi$ and $\vec{A}$ are the scalar and vector potentials, and $c$ is the speed of light. Also, 
\begin{gather}\label{eq:lp}
\mcc{L}_p = \sum_i \delta(\vec{x} - \vec{x}_i)\,L_i(\vec{x}, \vec{v}_i; \varphi, \vec{A}),
\end{gather}
where the summation is taken over individual particles, and $L_i$ are the Lagrangians of those particles; namely, $L_i = L_i^{(0)} + L_i^{(\rm int)}$, where $L_i^{(0)}$ are independent of the field, and $L_i^{(\rm int)} = (e_i/c) (\vec{v}_i \cdot \vec{A})- e_i \varphi$. Finally, $\vec{x}_i(t)$ are the trajectories of individual particles, $\vec{v}_i = \dot{\vec{x}}_i$ are the corresponding velocities, and $e_i$ are the particle charges.

Suppose that the electromagnetic field contains a rapidly oscillating part and consider the plasma dynamics on scales large compared to the oscillation scales. (In the presence of resonant or trapped particles, one of such scales is the period of bounce oscillations $\tau_b$ \cite[Sec.~8-6]{book:stix}.) Then, it is only the time-averaged part of the Lagrangian, $\msf{L}_\Sigma \equiv \int \favr{\mcc{L}_\Sigma}\,d\msf{V}$, that contributes to the system action. Hence $\msf{L}_\Sigma$ plays a role of the slow-motion Lagrangian of the system \cite{ref:whitham65}. Specifically, we write
\begin{gather}\label{eq:avrl1}
 \favr{\mcc{L}_\Sigma} = \favr{\mcc{L}_{\rm em}} + \favr{\mcc{L}^{(p)}_p} + \favr{\mcc{L}^{(t)}_p}.
\end{gather}
Here $\favr{\mcc{L}_{\rm em}}$ generally consists of two terms, $\bar{\mcc{L}}_{\rm em}$ due to quasistatic fields $(\bar{\varphi}, \bar{\vec{A}})$ (if any) and $\favr{\tilde{\mcc{L}}_{\rm em}}$ due to the actual wave field. The remaining terms describe contributions of passing particles and trapped particles, correspondingly, and are derived as follows.

In the case of passing particles, we separate the slow, OC motion $\bar{\vec{x}}_i(t)$ and the quiver motion $\tilde{\vec{x}}_i(t)$ and notice that
\begin{multline}
\int \favr{\delta (\vec{x} - \bar{\vec{x}}_i - \tilde{\vec{x}}_i)\, L^{(p)}_i(\vec{x}, \vec{v}_i)}\,d\msf{V} =\\
\int \favr{\delta (\vec{x} - \bar{\vec{x}}_i)\, L^{(p)}_i(\vec{x} + \tilde{\vec{x}}_i, \bar{\vec{v}}_i + \tilde{\vec{v}}_i)}\,d\msf{V} =\\
\int \delta (\vec{x} - \bar{\vec{x}}_i)\, \favr{L^{(p)}_i(\vec{x} + \tilde{\vec{x}}_i, \bar{\vec{v}}_i + \tilde{\vec{v}}_i)}\,d\msf{V} \equiv \\
\int \delta (\vec{x} - \bar{\vec{x}}_i)\, \mc{L}^{(p)}_i(\vec{x}, \bar{\vec{v}}_i)\,d\msf{V}.
\end{multline}
Here we introduced
\begin{gather}
\mc{L}^{(p)}_i(\vec{x}, \vec{v}) = \favr{L_i(\vec{x} + \tilde{\vec{x}}_i, \vec{v} + \tilde{\vec{v}}_i)},
\end{gather}
which has the meaning of a single-particle OC Lagrangian \cite{arX:mycoin}. Hence, one can write
\begin{gather}\label{eq:larg3}
 \favr{\mcc{L}^{(p)}_p} = \sum_i \delta (\vec{x} - \bar{\vec{x}}_i)\, \mc{L}^{(p)}_i(\vec{x}, \bar{\vec{v}}_i).
\end{gather}

In the case of trapped particles we proceed similarly, except that the OC location $\bar{\vec{x}}_i(t)$ is now determined by the motion of the wave nodes and thus cannot serve as an independent variable. Instead, the new independent variable will be the phase $\vec{\theta}_i$ of bounce oscillations, possibly in multiple dimensions. Since these bounce oscillations are assumed adiabatic, $\mc{L}^{(t)}_i$ will not depend on $\vec{\theta}_i$ explicitly; rather it will depend on $\dot{\vec{\theta}}_i$ and, parametrically, on $\bar{\vec{x}}_i$. (Remember that the dependence on the field variables is also implied throughout the paper.) Thus, the Lagrangian of the bounce motion, henceforth also called OC Lagrangian for brevity, is given by
\begin{gather}
\mc{L}^{(t)}_i(\vec{x}, \dot{\vec{\theta}}_i) = \favr{L_i(\vec{x} + \tilde{\vec{x}}_i, \vec{u} + \tilde{\vec{v}}_i)},
\end{gather}
where we substituted the wave phase velocity $\vec{u}$ for the average velocity. This yields
\begin{gather}
 \favr{\mcc{L}^{(t)}_p} = \sum_i \delta (\vec{x} - \bar{\vec{x}}_i)\, \mc{L}^{(t)}_i(\vec{x}, \vec{\theta}_i).
\end{gather}

\subsection{Routhian}

Below, it will be more convenient to use canonical OC variables for particles, $(\vec{q}_i, \vec{p}_i)$. For simplicity, let us temporarily require that $\vec{q}_i = \bar{\vec{x}}_i$ for passing particles and $\vec{q}_i = \vec{\theta}_i$ for trapped particles; in the latter case the canonical momentum $\vec{p}_i$ will be the action $\vec{J}_i$ of the bounce oscillations. Then, let us use \cite{foot:scalar}
\begin{gather}
 \mc{L}_i = \vec{p}_i \cdot \dot{\vec{q}}_i - \mc{H}_i,
\end{gather}
so \Eq{eq:avrl1} rewrites as
\begin{gather}\label{eq:lsigf}
\msf{L}_\Sigma = \msf{L} + \sum_i \vec{p}_i \cdot \dot{\vec{q}}_i,
\end{gather}
where $\msf{L} = \int \mcc{L}\,d\msf{V}$, and
\begin{gather}\label{eq:routh1}
\mcc{L} = \favr{\mcc{L}_{\rm em}} - \sum_i \delta (\vec{x} - \bar{\vec{x}}_i)\, \mc{H}_i(\vec{x}, \vec{p}_i).
\end{gather}
Since both $\vec{q}_i(t)$ and $\vec{p}_i(t)$ are now independent functions (cf. Ref.~\cite[Sec.~40]{book:landau1}), \textit{field} equations will be insensitive to the second term in \Eq{eq:lsigf}; \ie for the purpose of finding \textit{field} equations, this term can be dropped. Therefore, $\msf{L}$ plays the role of the adiabatic Lagrangian of the wave (and also of quasistatic fields, if any).

Notice that, since $\msf{L} = \msf{L}_\Sigma - \sum_i \vec{p}_i \cdot \dot{\vec{q}}_i$, it can be considered as a Routhian of the particle-field system \cite[Sec.~41]{book:landau1}, \ie a function that acts as a Lagrangian for the field variables but as a Hamiltonian for the particle variables. As a Routhian, the wave Lagrangian was also introduced earlier in our \Ref{my:bgk}. (However, unlike in \Ref{my:bgk}, here we do not perform Routh \textit{reduction} (\App{app:routh}); \ie now we allow $\vec{p}$ to evolve.) Below, we will show how \Eq{eq:routh1} corresponds to that earlier result.

\subsection{Locally averaged densities}

In \Eq{eq:routh1}, the summation over all particles $i$ can be separated into (i)~summation over species $s$, (ii)~summation over $\vec{p}_j$ within a local elementary spatial volume $\Delta\msf{V}_k$, and (iii)~summation over all $\Delta\msf{V}_k$. Specifically, let us choose the elementary volumes large enough such that both $\mc{H}_s$ and the densities $n_s$ vary little \cite{foot:avrn} within $\Delta\msf{V}_k$. Then,
\begin{align}
\sum_i & \delta (\vec{x} - \bar{\vec{x}}_i)\, \mc{H}_i(\vec{x}, \vec{p}_i) \notag\\
 & = \sum_s \sum_{k} \delta (\vec{x} - \bar{\vec{x}}_k)\, \sum_j \mc{H}_s (\bar{\vec{x}}_k, \vec{p}_j) \notag\\
 & = \sum_s \sum_{k} \delta (\vec{x} - \bar{\vec{x}}_k)\, n_s(\bar{\vec{x}}_k, t) \favr{\mc{H}_s(\bar{\vec{x}}_k, \vec{p})}_{\avp} \Delta\msf{V}_k \notag\\
 & = \sum_s n_s(\vec{x},t) \favr{\mc{H}_s(\vec{x}, \vec{p})}_{\avp}.
\end{align}
This puts $\mcc{L}$ in the anticipated form, \Eq{eq:lorig}, or
\begin{gather}\label{eq:finalL}
\mcc{L} = \favr{\mcc{L}_{\rm em}} - \sum_s n^{(p)}_s \favr{\mc{H}^{(p)}_s}_{\avp} - \sum_s n^{(t)}_s \favr{\mc{H}^{(t)}_s}_{\avp}.
\end{gather}
From now on, the specific canonical variables will not matter; \ie  further canonical transformations are allowed in $\mc{H}^{(p)}_s$ and $\mc{H}^{(t)}_s$, if necessary.

\subsection{Independent variables}

We are now to choose the independent variables that will describe the field. [Quasistatic fields, if any, can be described by $(\bar{\varphi}, \bar{\vec{A}})$ as usual and thus will not be considered explicitly.] Suppose that the wave is characterized by a smooth envelope $a(\vec{x}, t)$, arbitrarily normalized. Also suppose that the wave field, while not necessarily monochromatic, oscillates rapidly with some canonical phase $\xi$, the period being $2\pi$. Hence, the local temporal and spatial periods can be defined as $T = 2\pi/\omega$ and $\lambda = 2\pi/k$, such that $u = \omega/k$ is the phase speed, $\vec{u} = u\vec{k}/k$,
\begin{gather}\label{eq:omk}
\omega = - \pd_t \xi, \quad \vec{k} = \nabla \xi,
\end{gather}
and, in particular,
\begin{gather}\label{eq:consistency}
\pd_t k_i + \pd_i \omega = 0, \quad \pd_j {k_i} - \pd_i {k_j} = 0.
\end{gather}
where we introduced $\pd_i \equiv \pd_{x_i}$. The function $\mcc{L}$ will then depend on $(\pd_t\xi, \nabla \xi)$ \cite{foot:xt} but not on $\xi$, for it describes the dynamics on scales $\tau$ and $\lambda$ such that $\tau \gg T$ and $\ell \gg \lambda$~\cite{ref:whitham65}. (In the presence of trapped particles, we also require $\tau \gg \tau_b$ and $\ell \gg u\tau_b$.)

The question that remains is how to treat $n_s$ when varying $\mcc{L}$. For passing particles, the OC densities $n^{(p)}_s$ are determined by $\bar{\vec{x}}_i = \vec{q}_i$, which are independent variables; thus, $n^{(p)}_s (\vec{x},t)$ are also independent of the field variables. For trapped particles, however, $n^{(t)}_s (\vec{x},t)$ are determined by $\bar{\vec{x}}_i$ which may be tied to the wave troughs; thus, $n^{(t)}_s (\vec{x},t)$ may be connected with the wave phase. In particular, for 1D waves this connection can be implemented as an exact closure, which is done as follows.

\section{One-dimensional waves with trapped particles}
\label{sec:1dtrap}

\subsection{Extended Lagrangian}

First of all, notice that, in a 1D system, trapped particles travel at the wave phase velocity $u$. Hence, the corresponding continuity equations read as
\begin{gather}\label{eq:cont}
\pd_t \mcc{n}_s + \pd_x (\mcc{n}_s u) = 0,
\end{gather}
where we introduced $\mcc{n}_s \equiv n^{(t)}_s$ to shorten the notation. One can embed \Eq{eq:cont} in the formalism by considering a new, extended Lagrangian density
\begin{gather}\label{eq:Lambda}
\Lambda = \mcc{L} + \sum_s \mu_s \big[\pd_t \mcc{n}_s + \pd_x (\mcc{n}_s u) \big].
\end{gather}
Here $\mu_s$ are Lagrange multipliers \cite{ref:seliger68}, \ie new independent functions of $(t, x)$, yet to be found. In particular, varying $\Lambda$ with respect to $\mcc{n}_s$ yields
\begin{gather}\label{eq:mucont}
\pd_t \mu_s + u\, \pd_x \mu_s + \favr{\mc{H}^{(t)}_s}_{\avp} = 0,
\end{gather}
whereas \Eq{eq:cont} flows from varying $\Lambda$ with respect to~$\mu_s$. 

Further, notice that
\begin{gather}\label{eq:Lambda2}
\hat{\Lambda} = \mcc{L} - \sum_s \mcc{n}_s \big[\pd_t \mu_s + (\omega/k)\,\pd_x \mu_s \big]
\end{gather}
is a Lagrangian density equivalent to $\Lambda$, yet with an advantage that $\hat{\Lambda}$ depends on only \cite{foot:xt} the \textit{first}-order derivatives of $\xi$ [cf. \Eq{eq:omk}]:
\begin{gather}\label{eq:Lambda3}
\hat{\Lambda} = \hat{\Lambda} (a, \pd_t \xi, \pd_x \xi, \mcc{n}, \pd_t \mu, \pd_x \mu).
\end{gather}
Then, varying $\hat{\Lambda}$ with respect to $\xi$ is as usual and yields $\pd_t \hat{\Lambda}_\omega - \pd_x \hat{\Lambda}_k = 0$ \cite{ref:whitham65}. (We henceforth use indexes $\omega$, $k$, and $a$ to denote the corresponding partial derivatives.) On the other hand, 
\begin{gather}
\hat{\Lambda}_\omega = \mcc{L}_\omega - \frac{1}{k} \sum_s \mcc{n}_s\,\pd_x\mu_s, \\
\hat{\Lambda}_k = \mcc{L}_k + \frac{\omega}{k^2} \sum_s \mcc{n}_s\,\pd_x\mu_s,
\end{gather}
where the derivatives are taken, in particular, at fixed $\mcc{n}$. Thus, one obtains
\begin{gather}\label{eq:act1}
\pd_t \mcc{L}_\omega - \pd_x \mcc{L}_k = \sum_s M_s, \\
M_s = \pd_t(\sigma_s\, \pd_x \mu_s) + \pd_x (\sigma_s u\,\pd_x\mu_s).
\end{gather}
Here we introduced $\sigma_s = \mcc{n}_s/k$, which is proportional to the number of trapped particles (of type $s$) within one wavelength. For adiabatic waves this number is constant in the frame moving with the phase velocity; \ie
\begin{gather}\label{eq:sigmaeq}
\pd_t \sigma_s + u \,\pd_x \sigma_s = 0,
\end{gather}
which is also obtained from \Eqs{eq:consistency} and \eq{eq:cont}. (Notice that the equation for $\sigma_s$ does \textit{not} have a form of a continuity equation, unlike that for $\mcc{n}_s$.) Then,
\begin{gather}
 M_s = \sigma_s \big[\pd^2_{xt}\mu_s + \pd_x (u\,\pd_x\mu_s) \big] = - \sigma_s\,  \pd_x \favr{\mc{H}^{(t)}_s}_{\avp},
\end{gather}
where we used \Eq{eq:mucont}. Hence, one gets
\begin{gather}\label{eq:act2}
\pd_t \mcc{L}_\omega - \pd_x \mcc{L}_k = - \sum_s \sigma_s\,  \pd_x \favr{\mc{H}^{(t)}_s}_{\avp},
\end{gather}
or, equivalently,
\begin{multline}\label{eq:act3}
\pd_t \mcc{L}_\omega + \pd_x \Big[ - \mcc{L}_k + \sum_s \sigma_s \favr{\mc{H}^{(t)}_s}_{\avp} \Big] = \\
\sum_s \favr{\mc{H}^{(t)}_s}_{\avp}\, \pd_x \sigma_s.
\end{multline}
Further notice that $(\mcc{L}_k)_{\sigma} = (\mcc{L}_k)_{\mcc{n}} + \sum_k \sigma_s (\pd_{\mcc{n}_s} \mcc{L})_k$, so
\begin{gather}\label{eq:sigmaton}
(\mcc{L}_k)_{\sigma} = (\mcc{L}_k)_{\mcc{n}} - \sum_k \sigma_s \favr{\mc{H}^{(t)}_s}_{\avp},
\end{gather}
where we substituted $\mcc{n}_s = \sigma_s k$, in the left-hand side; similarly, $(\mcc{L}_\omega)_\sigma = (\mcc{L}_\omega)_{\mcc{n}}$. (Here the external subindexes show variables kept fixed at differentiation.) Thus, it is convenient to consider $\mcc{L}$ as a function of $\sigma_s$ rather than of $\mcc{n}_s$, specifically as follows.

\subsection{Action conservation and wave dispersion}

From now on, let us consider $\mcc{L}$ as \cite{foot:xt}
\begin{gather}\label{eq:lsigma}
\mcc{L} = \mcc{L}(a, \pd_t \xi, \pd_x\xi, \sigma).
\end{gather}
Using \Eq{eq:sigmaton}, one can hence write \Eq{eq:act3} as
\begin{gather}\label{eq:act4}
\pd_t \mcc{L}_\omega - \pd_x \mcc{L}_k = \sum_s \favr{\mc{H}^{(t)}_s}_{\avp}\, \pd_x \sigma_s.
\end{gather}
Equation \eq{eq:act4} represents a generalization of the well-known ACT for 1D waves \cite{ref:whitham65}, reproduced in the limit $\sigma_s = 0$ (also see \App{app:lin}). Thus, we interpret
\begin{gather}\label{eq:mcidef}
\mc{I} = \mcc{L}_\omega, \quad  \mc{J} = - \mcc{L}_k
\end{gather}
as the new action density and the new action flux density, correspondingly. Notice, however, that \Eq{eq:act4}, or
\begin{gather}\label{eq:act1dmain}
\pd_t \mc{I} + \pd_x \mc{J} = \sum_s \favr{\mc{H}^{(t)}_s}_{\avp}\, \pd_x \sigma_s,
\end{gather}
does not have a conservative form in general, due to the nonzero right-hand side. This is because the wave of the trapped-particle density is, by definition, always resonant with the electric field, so the two can exchange quanta whenever $\sigma_s$ are modulated. Interestingly, the effect of $\sigma$-waves, which are described by \Eq{eq:sigmaeq}, is similar to the effect of entropy waves on magnetohydrodynamic oscillations reported in \Refs{ref:webb05, ref:webb07}.

Finally, we can complement the ACT with the NDR, by varying $\mcc{L}$ with respect to the wave amplitude. Withing the lowest-order (in $\tau^{-1}$ and $\ell^{-1}$) GO approximation, to which we adhere throughout the paper, $\mcc{L}$ depends on the local amplitude $a$ but not on its derivatives [\Eq{eq:lsigma}], as usual \cite{ref:whitham65, foot:high}. Remember also that $\sigma_s$ are introduced as independent functions and thus do not depend on $a$ by definition. Hence, the NDR attains the same general form as for adiabatic waves without trapped particles, namely,
\begin{gather}\label{eq:genndr}
\mcc{L}_a = 0.
\end{gather}

\section{Longitudinal waves}
\label{sec:lwaves}

Now let us consider the specific case of 1D longitudinal waves as a paradigmatic example. To do so, we will need to construct the Lagrangian density $\mcc{L}$ for such waves, which, in turn, requires calculating the single-particle OC Hamiltonians $\mc{H}_s$ first. For linear waves, this is done in \App{app:lin}. For nonlinear waves, this also can be done straightforwardly, at least to the zeroth order in $\tau^{-1}$ and~$\ell^{-1}$. Namely, we proceed as follows.

\subsection{Single-particle OC Hamiltonians}

Consider the Lagrangian of a single particle in a stationary homogeneous electrostatic wave in nonmagnetized plasma. It is only the 1D motion along the wave field $\tilde{\vec{E}}$ that matters for us; thus, we take
\begin{gather}
L = mv^2/2 - e\tilde{\varphi}(x - ut),
\end{gather}
where $m$ and $e$ are the particle mass and charge, $v$ is the velocity in the laboratory frame $K$, and the potential $\tilde{\varphi}$ is periodic yet not necessarily sinusoidal. (A quasistatic potential $\bar{\varphi}$ can be included straightforwardly and will not be discussed here explicitly.) Rewrite $L$ as
\begin{gather}
L = mu^2/2 + m u w + mw^2 - \mc{E},
\end{gather}
where $y = x - ut$, so $\dot{y} \equiv w = v - u$, and also
\begin{gather}
\mc{E} \equiv mw^2/2 + e\tilde{\varphi}(y),
\end{gather}
which is the energy in the moving frame $\hat{K}$ where $\tilde{\varphi}$ is static. Then, time-averaging yields
\begin{gather}
\favr{L} = mu^2/2 + m u \favr{w} + m\favr{w^2} - \mc{E},
\end{gather}
where we used that $\mc{E}$ is conserved on the oscillation scale.

To proceed, it is convenient to introduce the angle $\theta$ and the action $J$ of the oscillations in $\hat{K}$; in partucular,
\begin{gather}
J = \frac{m}{2\pi}\oint w\,dy.
\end{gather}
Then the oscillation period can be expressed in terms of the corresponding canonical frequency $\Omega$, yielding $\favr{w} = \varsigma \Omega/k$, where $\varsigma \equiv \text{sgn}\, \favr{w} = \text{sgn}\, w$ (for trapped particles $\varsigma = 0$), and $m\favr{w^2} = J\Omega$. Then,
\begin{gather}
\favr{L} = muV - mu^2/2 + \pounds,
\end{gather}
where we used that the average velocity $V \equiv \favr{v}$ equals
\begin{gather}\label{eq:V}
V = u + \varsigma \Omega/k
\end{gather}
and introduced $\pounds = J\Omega - \mc{E}$. 

In particular, notice the following. Since the generating function of the transformation $(y, mw) \to (\theta, J)$ clearly does not depend on time explicitly, $\mc{E}$ acts as a Hamiltonian in $(\theta, J)$-representation (cf. \Ref{my:bgk}), and thus $\pounds$ is the corresponding Lagrangian; then, 
\begin{gather}\label{eq:eom}
\pd_J \mc{E} = \Omega, \quad \pd_J \pounds = J\pd_J \Omega.
\end{gather}
[Partial derivatives are used because $\mc{E}$ and $\pounds$ can also depend parametrically on $(a, \omega, k)$; cf. \Sec{sec:iv}.]

First, let us consider a passing particle. In this case, for the canonical momentum $\vec{p}$ one can take $P \equiv \pd_V \favr{L}$ \cite{arX:mycoin, my:mneg}. Assuming $J = J(V)$, one thereby obtains $P = mu + \pd_J \pounds\, \pd_V J$, or, using \Eq{eq:eom}, $P = mu + J\,\pd_{V} \Omega$. Hence, from \Eq{eq:V}, we get
\begin{gather}\label{eq:PJ}
P = mu + \varsigma k J.
\end{gather}
Then the OC Hamiltonian $\mc{H} = PV - \favr{L}$ reads as
\begin{gather}\label{eq:hampass}
\mc{H} = \mc{E} + Pu - mu^2/2.
\end{gather}
In particular, notice that \Eq{eq:hampass} can be understood as a generalization of the nonrelativistic dipole ponderomotive Hamiltonian [\Eq{eq:ponder}, with $\Phi_s$ from \Eq{eq:lponder}] to the case of fully nonlinear particle motion in an arbitrary longitudinal electrostatic wave.

In case of a trapped particle, the average coordinate is fixed, yielding $V = u$, so now those are $(\theta, J)$ that we choose to serve as $(\vec{q}, \vec{p})$. Hence, $\mc{H} = J\Omega - \favr{L}$, and thus
\begin{gather}\label{eq:hamtr}
\mc{H} = \mc{E} - mu^2/2.
\end{gather}

\subsection{Parametrization. Wave Lagrangian}
\label{sec:iv}

Although $a$ can be defined as an arbitrary measure of the field amplitude, for the purpose of this paper it is convenient to introduce it specifically as the amplitude of the wave electric field $\tilde{E}$. Hence, we can write \Eq{eq:finalL} explicitly as
\begin{gather}\label{eq:lwmainl}
\mcc{L} = \bar{\mcc{L}}_{\rm em} + \frac{a^2}{16\pi} - \sum_s n^{(p)}_s \favr{\mc{H}_s^{(p)}}_{\avp} - \sum_s n_s^{(t)} \favr{\mc{H}_s^{(t)}}_{\avp},
\end{gather}
with $\mc{H}_s^{(p)}$ to be taken from \Eq{eq:hampass}, and $\mc{H}^{(t)}_s$ to be taken from \Eq{eq:hamtr}.

Notice also that, once we have adopted $a \equiv \tilde{E}$, the bounce-motion energy $\mc{E}$ can depend parametrically on $a$ and $k$ but not on $\omega$, because the particle motion in $\hat{K}$ is entirely determined by the spatial structure of the wave potential, which is static there. In other words,
\begin{gather}\label{eq:mce2}
\mc{E} = \mc{E}(J, a, k).
\end{gather}
Yet note that $J = J(P, \omega, k)$ for passing particles [see \Eq{eq:PJ}], whereas for trapped particles $J$ is an independent variable. In particular, this yields the following equalities that we will use below. First of all,
\begin{gather}
\pd_\omega \mc{E}^{(p)} = \pd_J \mc{E}^{(p)}\,\pd_\omega J^{(p)} = - m\varsigma\Omega/k^2,
\end{gather}
where we substituted \Eq{eq:eom} for $\pd_J \mc{E}$ and \Eq{eq:PJ} for $\pd_\omega J(P, \omega, k)$; also, $\pd_\omega \mc{E}^{(t)} = 0$. Hence, one obtains
\begin{gather}\label{eq:hw}
\pd_\omega \mc{H}^{(p)} = (P - mV)/k, \quad \pd_\omega \mc{H}^{(t)} = - mu/k.
\end{gather}

\subsection{Action density}
\label{sec:actd}

Now we can calculate the wave action density $\mc{I}$ [\Eq{eq:mcidef}], namely, as follows. Since the first two terms in \Eq{eq:lwmainl} are independent of $\omega$ and $k$, one gets
\begin{gather}\label{eq:act21}
\mc{I} = \mc{I}^{(p)} + \mc{I}^{(t)}, \quad \mc{I}^{(b)} = - \sum_s n^{(b)}_s \favr{\pd_\omega \mc{H}_s^{(b)}}_{\avp},
\end{gather}
with $b = p, t$. Hence, \Eq{eq:hw} yields
\begin{gather}
\mc{I}^{(p)} = k^{-1}\sum_s n_s^{(p)}\, \favr{m_s V - P}_{\avp}, \label{eq:ipt}\\ 
\mc{I}^{(t)} = k^{-1}\sum_s n_s^{(t)} m_s u.\label{eq:ipt2}
\end{gather}

One may say that $\mc{I}^{(p)}$ is proportional to (the density of) the \textit{ponderomotive} momentum carried by passing particles (cf. \App{app:lew}), and $\mc{I}^{(t)}$ is proportional to the \textit{kinetic} momentum carried by trapped particles. Also, notice that the two can be combined~as
\begin{gather}\label{eq:wp}
 k\mc{I} = \sum_s n_s m_s \favr{V}_{\avp} - \sum_s n_s^{(p)} \favr{P}_{\avp}.
\end{gather}
The right-hand side here equals the difference between the system total kinetic momentum [$n_s \equiv n_s^{(p)} + n_s^{(t)}$ being the total density of species $s$] less the momentum stored in particles, \ie that of the untrapped population. Therefore, by definition, $k\mc{I}$ represents the wave total momentum, in agreement with Ref.~\cite[Sec.~15.4]{book:whitham} (see also Paper~III). Notice that a part of this momentum [namely, $k\mc{I}^{(t)}$] is independent of $a$, because it is stored in the trapped-particle translational motion with velocity $u = \omega/k$. However, remember that we still assume that $a$ must remain large enough, such that detrapping does not occur (\Sec{sec:intro}).

\subsection{Action flow}

The action flux density $\mc{J}$ [\Eq{eq:mcidef}] can be found similarly from \Eq{eq:lwmainl} and reads as $\mc{J} = \mc{J}^{(p)} + \mc{J}^{(t)}$, where
\begin{gather}\label{eq:mcj2}
\mc{J}^{(b)} = \sum_s n_s^{(b)}\,\pd_k \favr{\mc{H}_s^{(b)}}_{\avp},
\end{gather}
again with $b = p, t$. In particular, notice that a part of $\mc{J}$ is also independent of $a$, like $\mc{I}^{(t)}$. This is because $\mc{J} = \Pi/\omega$, where $\Pi$ is the energy flux density \cite[Sec.~15.4]{book:whitham} (see also Paper~III), a part of which is determined by the trapped-particle average velocity $u$ rather than $a$. Notice also that $v_{\mc{I}} \equiv \mc{J}/\mc{I}$ has the meaning of the action flow velocity and reads as
\begin{gather}
v_{\mc{I}} = \frac{\mc{J}^{(p)} + \mc{J}^{(t)}}{\mc{I}^{(p)} + \mc{I}^{(t)}} .
\end{gather}
In particular, with $\varrho \equiv \mc{I}^{(t)}/\mc{I}^{(p)}$ and $v_{\mc{I}}^{(b)} = \mc{J}^{(b)}/\mc{I}^{(b)}$, one gets
\begin{gather}\label{eq:vmci2}
v_{\mc{I}} = \frac{v_{\mc{I}}^{(p)} + \varrho v_{\mc{I}}^{(t)}}{1 + \varrho}.
\end{gather}

\section{Discussion}
\label{sec:discuss}

Now that we have developed the general formalism, it is instructive to consider 1D longitudinal electrostatic waves at small $a$ in particular. Then, one can use \Eq{eq:ponder} for $\mc{H}^{(p)}_s$ \cite{foot:notenough}, so
\begin{gather}\label{eq:lwmainlin}
\mcc{L} \approx \mcc{L}^{(0)} + \frac{\epsilon a^2}{16\pi} - \sum_s n_s^{(t)} \favr{\mc{H}_s^{(t)}}_{\avp},
\end{gather}
where $\mcc{L}^{(0)}$ is independent of the field variables, and $\epsilon$ is the longitudinal dielectric function (\App{app:lin}). Also, $\mc{I}^{(p)} \approx \epsilon_\omega |\tilde{E}|^2/(16\pi)$, and thus $v_{\mc{I}}^{(p)}$ equals the linear group velocity $v_{g0} = \omega_k$ (\App{app:lew}). Further, let us neglect $\mc{E}_s^{(t)}$ compared to $m_s u^2/2$ in \Eq{eq:hamtr}. Then, $\mc{H}_s^{(t)} \approx - m_s\omega^2/(2k^2)$, so one obtains
\begin{gather}
\mc{J}^{(t)} \approx \sum_s n_s^{(t)}\,m_s u^2/k = u \mc{I}^{(t)},
\end{gather}
which is independent of $a$, as expected. This yields ${v_{\mc{I}}^{(t)} \approx u}$, and, therefore,
\begin{gather}\label{eq:vglin}
v_{\mc{I}} \approx \frac{v_{g0} + \varrho u}{1 + \varrho}.
\end{gather}

Equation \eq{eq:vglin} should not be confused with a similar expression in \Ref{ref:benisti10} derived for what is called there the nonlinear group velocity. The effects addressed in \Ref{ref:benisti10} result in a nonconservative form of the envelope equation, \ie violate the ACT; hence, they are dissipative by definition. In contrast, our formulation does not account for collisionless dissipation (except at inhomogeneous $\sigma_s$); thus, in \Eq{eq:vglin} the difference between $v_{\mc{I}}$ and $v_{g0}$ is entirely due to adiabatic effects \cite{foot:landau}. 

In addition to $v_{\mc{I}}$ that we presented, one can also define the velocities of the energy and momentum flows \cite{ref:whitham65}. For a nonlinear wave, those will be different from each other and from the true nonlinear group velocities $v_g$, at which the modulation impressed on a wave propagates adiabatically \cite{foot:impressed}. The effect of trapped particles on those true $v_g$ will be discussed in Paper~III.

\section{Summary}
\label{sec:concl}

In this paper, a Lagrangian formalism is developed for general nondissipative quasiperiodic nonlinear waves in collisionless plasma. Specifically, the time-averaged adiabatic Lagrangian density is derived in the following form:
\begin{gather}
\mcc{L} = \favr{\mcc{L}_{\rm em}} - \sum_s n_s \favr{\mc{H}_s}_{\avp}.
\end{gather}
Here $\favr{\mcc{L}_{\rm em}}$ is the time-averaged Lagrangian density of the electromagnetic field, summation is taken over distinct species $s$, $n_s$ are the corresponding average densities, and $\favr{\mc{H}_s}_{\avp}$ are the corresponding OC Hamiltonians averaged over the distributions $f_s$ of canonical momenta. Once $\mc{H}_s$ are found, the complete set of GO equations is derived without referring to the Maxwell-Vlasov system.

For the first time, the average Lagrangian accounts also for particles trapped by the wave, under the assumption that the number of these particles remains fixed (\Sec{sec:intro}). In particular, 1D waves are considered, in which case $\mcc{L} = \mcc{L}(a, \omega, k, \sigma)$; here $\omega$ and $k$ are the wave local frequency and the wave number, and $\sigma_s \equiv n_s^{(t)}/k$ are proportional to the number of trapped particles within a wavelength. Correspondingly, the GO equations are summarized as follows. The first one is the consistency condition, $\pd_t k + \pd_x \omega = 0$, due to $\omega = - \pd_t \xi$ and $k = \pd_x \xi$; here $\xi$ is the wave canonical phase. The second one is the NDR, given by $\mcc{L}_a = 0$. The third GO equation is a modified ACT,
\begin{gather}
\pd_t \mc{I} + \pd_x \mc{J} = \sum_s \mc{H}_s^{(t)}\, \pd_x \sigma_s,
\end{gather}
with $\mc{I} \equiv \mcc{L}_\omega$ and $\mc{J} \equiv - \mcc{L}_k$ being the action density and the action flux density, correspondingly. Because of the source term on the right-hand side, the wave action may not be conserved then, due to the fact that it can be exchanged with resonant waves of the trapped-particle density ($\sigma$-waves). 

Since the number of trapped particles is fixed within our model, their contributions to the momentum density $\rho$ and the energy flux density $\Pi$ depend mainly on $\sigma_s$ and the phase velocity $u$, rather than on the wave amplitude $a$ explicitly. Hence, $\mc{I}$ and $\mc{J}$ (and thus $\mcc{L}$ too) may contain $a$-independent terms, because
\begin{gather}
\mc{I} = \rho/k, \quad \mc{J} = \Pi/\omega.
\end{gather}
Particularly, in the limit of small $a$, the action flow velocity is obtained,
\begin{gather}\label{eq:vi}
v_{\mc{I}} \approx \frac{v_{g0} + \varrho u}{1 + \varrho},
\end{gather}
where $v_{g0} = \omega_k$ is the linear group velocity, and $\varrho = \mc{I}^{(t)}/\mc{I}^{(p)}$. The difference between $v_{\mc{I}}$ and $v_{g0}$ here is entirely due to adiabatic effects, so \Eq{eq:vi} should not be confused with a seemingly akin formula in \Ref{ref:benisti10}.

Applications of these results are left to Papers II and III, where specific problems are addressed pertaining to properties and dynamics of waves with trapped particles.

\section{Acknowledgments}

The work was supported through the NNSA SSAA Program through DOE Research Grant No. DE274-FG52-08NA28553.

\appendix

\section{Routh reduction}
\label{app:routh}

In this appendix, we restate the concept of Routh reduction \cite{book:arnold06}, complementing the derivation of the wave Lagrangian that we reported earlier in \Ref{my:bgk}. 

Consider a dynamical system described by generalized coordinates $\vec{q} \equiv (q^1, \ldots q^N)$, so the corresponding Lagrangian has a form $L(\vec{q}, \dot{\vec{q}}, t)$, $t$ being the time. The original least-action principle is then formulated as follows \cite[Sec.~2]{book:landau1}: among trajectories $\vec{q}(t)$ starting at $\vec{q}_1$ at time $t_1$ and ending at $\vec{q}_2$ at time $t_2$, realized is the one on which the action $S = \int^{t_2}_{t_1} L\,dt$ is minimal. Since the general variation of $S$ reads~as \cite{foot:scalar}
\begin{gather}\label{eq:dsdef}
\delta S = \int^{t_2}_{t_1} \left(\frac{\partial L}{\partial \vec{q}}-\frac{d}{dt}\,\frac{\partial L}{\partial \dot{\vec{q}}}\right)\cdot\delta \vec{q}\,dt+\vec{p}\cdot\delta\vec{q}\,\Big|_{t_1}^{t_2},
\end{gather}
with $\vec{p} \equiv \pd_{\dot{\vec{q}}}L$, and $\delta \vec{q}(t_{1,2}) = 0$ due to $\vec{q}_{1,2}$ being fixed, one thereby obtains the Euler-Lagrange equations
\begin{gather}\label{eq:eul}
\frac{d}{dt}\,\frac{\partial L}{\partial \dot{\vec{q}}} = \frac{\partial L}{\partial \vec{q}}.
\end{gather}

For simplicity, we henceforth consider a 2D system, with $\vec{q} = (\theta, x)$, the extension to a larger number of dimensions being straightforward. Suppose, in particular, that $\theta$ is cyclic, \ie does not enter $L$ explicitly. Then the momentum $J$ canonically conjugate to $\theta$ is conserved, so we can use the equality $J = \pd_{\dot{\theta}} L$ to express $\dot{\theta}$~as
\begin{gather}\label{eq:dq}
\dot{\theta} \equiv \Omega(J, x, \dot{x}, t).
\end{gather}
Hence, $S$ can be understood as a functional of $x(t)$ only.

Now consider the set of trajectories $x(t)$ starting at given $x_1$ at time $t_1$ and ending at given $x_2$ at time $t_2$, while $\theta_1$ and $\theta_2$ are arbitrary [albeit connected through \Eq{eq:dq}]. Suppose that $\bar{x}(t)$ satisfies \Eq{eq:eul} and consider the linear variation of $S[x(t)]$ with respect to $\delta x(t) = x(t) - \bar{x}(t)$. Then, from \Eq{eq:dsdef}, one obtains 
\begin{gather}\label{eq:ds2}
\delta S = J\,\delta\theta\,\Big|_{t_1}^{t_2}.
\end{gather}
Since $S$ is thereby \textit{not} minimized on $\bar{x}(t)$ within these variation procedure, consider another, ``reduced'' action
\begin{gather}\label{eq:ss0}
\hat{S} = S - \int^{t_2}_{t_1} J\dot{\theta}\,dt.
\end{gather}
Here, the latter term can also be put as $\int^{\theta_2}_{\theta_1} J\,d\theta$, where depending on $x(t)$ are only the integration limits. Therefore, its variation around $\bar{x}(t)$ equals $J\,\delta\theta\,|_{t_1}^{t_2}$, thus yielding ${\delta \hat{S}[\bar{x}(t)] = 0}$. Then a new variational principle can be formulated as follows: among trajectories $x(t)$ starting at $x_1$ at $t_1$ and ending at $x_2$ at $t_2$, \textit{with arbitrary} $\theta_1$ \textit{and} $\theta_2$, realized is $\bar{x}(t)$ on which $\hat{S}$ is minimal. 

Notice further that \Eq{eq:ss0} rewrites as $\hat{S}= \int^{t_2}_{t_1} R\,dt$, with the equivalent Lagrangian 
\begin{gather}\label{eq:l2}
R (x, \dot{x}, t) = L(x, \dot{x}, t, \Omega) - J\Omega
\end{gather}
[where $\Omega = \Omega(J, x, \dot{x}, t)$], also known as Routhian. Hence, the motion equation that flows from the new variational principle reads~as
\begin{gather}\label{eq:eu2}
\frac{d}{dt}\,\frac{\partial R}{\partial \dot{x}} = \frac{\partial R}{\partial x},
\end{gather}
also in agreement with the general Routh equations \cite[Sec.~41]{book:landau1}. Since $R$ is independent of $\theta$, the system phase space is effectively reduced now, and $x$-motion decouples, which is what constitutes the Routh reduction. For plasma physics applications of this technique, see \Refs{arX:mycoin, my:mneg, my:kchi, my:nlinphi, ref:larsson86, ref:pfirsch04}.

\section{Linear waves}
\label{app:lin}

Here, we show how the known GO equations for linear electromagnetic waves [which have $n^{(t)} = 0$] follow from the general Lagrangian formalism discussed in \Sec{sec:wavelagr}. 

\subsection{General electromagnetic waves}
\label{app:gew}

Let us take $\tilde{\vec{E}}, \tilde{\vec{B}} \propto e^{-i\omega t + i\vec{k}\cdot \vec{x}}$, and employ the dipole approximation for $\mc{H}_s$ \cite{my:kchi, my:nlinphi}, namely,
\begin{gather}\label{eq:ponder}
\mc{H}_s = \mc{H}_s^{(0)} + \Phi_s, \quad \Phi_s = -\tilde{\vec{E}}^*\cdot\oper{\alpha}_s\cdot\tilde{\vec{E}}/4,
\end{gather}
where $\mc{H}_s^{(0)}$ is some function of the particle canonical momenta, $\Phi_s$ is the ponderomotive potential, and $\oper{\alpha}_s$ is the linear polarizability. Then, since 
\begin{gather}\label{eq:eps}
 1 + 4\pi \sum_s n_s \favr{\oper{\alpha}_s}_{\avp} = \oper{\epsilon}(\omega, \vec{k}),
\end{gather}
where $\oper{\epsilon}$ is the linear dielectric tensor, one obtains
\begin{gather}
\mcc{L} = \mcc{L}^{(0)} + \frac{1}{16\pi}\,\left(\tilde{\vec{E}}^* \cdot \oper{\epsilon} \cdot \tilde{\vec{E}} - |\tilde{\vec{B}}|^2\right),
\end{gather}
where the term
\begin{gather}
\mcc{L}^{(0)} = \bar{\mcc{L}}_{\rm em} - \sum_s n_s \favr{\mc{H}_s^{(0)}}_{\avp}
\end{gather}
is independent of the wave variables. Now let us introduce the wave amplitude $a$ via $\tilde{\vec{E}} = a\vec{e}$ for the electric field envelope, where $\vec{e}$ determines polarization; hence, $\tilde{B} = |\vec{n} \times \vec{e}| a$, where $\vec{n} \equiv c\vec{k}/\omega$, and $c$ is the speed of light. Then,
\begin{gather}
\mcc{L} = \mcc{L}^{(0)} + \frac{a^2}{16\pi}\, \mcc{D}(\omega, \vec{k}),
\end{gather}
where we introduced
\begin{gather}
\mcc{D}(\omega, \vec{k}) = \vec{e}^* \cdot \oper{\epsilon} \cdot \vec{e} - |\vec{n} \times \vec{e}|^2.
\end{gather}

Varying the wave Lagrangian with respect to the amplitude $a$ yields the dispersion relation $\mcc{L}_a = 0$, or
\begin{gather}\label{eq:ldre}
\mcc{D}(\omega, \vec{k}) = 0,
\end{gather}
which coincides with the known dispersion relation at prescribed $\vec{e}$ \cite[Sec.~1-3]{book:stix}. In fact, the \textit{vector} equation,
\begin{gather}\label{eq:drlinvec}
\oper{\epsilon} \cdot \vec{e} + \vec{n} \times (\vec{n} \times \vec{e}) = 0,
\end{gather}
also can be recovered, namely, by varying $\mcc{L}$ with respect to $\vec{e}^*$. [One could, of course, vary $\mcc{L}$ also with respect to $\tilde{\vec{E}}^*$ and get \Eq{eq:drlinvec} immediately.]

Now let us vary $\mcc{L}$ with respect to the wave phase $\xi$, with \Eq{eq:omk} taken into account. Like for any other Lagrangian density of the form $\mcc{L}(a, \omega , \vec{k})$, one obtains then \cite{ref:whitham65}
\begin{gather}\label{eq:linact}
\pd_t \mcc{L}_\omega - \nabla \cdot \mcc{L}_{\vec{k}} = 0.
\end{gather}
The quantity $\mc{I} \equiv \mcc{L}_\omega$ can be written as
\begin{align}
\mc{I} 
 & = \frac{a^2}{16\pi}\left(\vec{e}^* \cdot \oper{\epsilon}_\omega \cdot \vec{e}\right) + \frac{a^2}{8\pi \omega}\,|\vec{n} \times \vec{e}|^2  \notag\\
 & = \frac{a^2}{16\pi \omega}
     \left[\vec{e}^* \cdot (\omega \oper{\epsilon}_\omega)\cdot \vec{e} + \vec{e}^* \cdot \oper{\epsilon} \cdot \vec{e} + |\vec{n} \times \vec{e}|^2\right] \notag\\
 & = \frac{1}{16\pi \omega}\,\left[\tilde{\vec{E}}^* \cdot \pd_\omega (\oper{\epsilon} \omega)\cdot \tilde{\vec{E}} + |\tilde{\vec{B}}|^2\right],
\end{align}
or $\mc{I} = \varepsilon/\omega$, where we used \Eq{eq:ldre} and introduced $\varepsilon$ for the linear-wave energy density \cite{my:kchi}; thus, $\mc{I}$ equals the linear-wave action density. Also, $\vec{\mc{J}} \equiv - \mcc{L}_{\vec{k}}$ can be put~as
\begin{gather}\label{eq:actfluxlin}
\vec{\mc{J}} 
 = - \frac{a^2}{16\pi}\, \mcc{D}_{\vec{k}} 
 = \frac{a^2}{16\pi}\,\mcc{D}_\omega\left(-\frac{\mcc{D}_{\vec{k}}}{\mcc{D}_\omega}\right)
 =  \vec{v}_{g0}\mc{I},
\end{gather}
where we used $- \pd_{\vec{k}} \mcc{D}/\pd_\omega \mcc{D} = \omega_{\vec{k}}$ [from \Eq{eq:ldre}], the latter being the linear group velocity $\vec{v}_{g0}$; thus, $\vec{\mc{J}}$ is the linear-wave action flux density. Hence \Eq{eq:linact} rewrites~as
\begin{gather}\label{eq:linact2}
\pd_t \mc{I} + \nabla \cdot (\vec{v}_{g0}\mc{I}) = 0,
\end{gather}
in agreement with the linear ACT \cite[Sec.~11.7]{book:whitham}.

Equations \eq{eq:omk}, \eq{eq:ldre}, and \eq{eq:linact2} represent a complete set of equations describing nondissipative linear electromagnetic waves in the GO approximation (cf. Ref.~\cite[Chaps.~14, 15]{book:whitham}). As one can see from the above calculation, the Maxwell's equations and the Vlasov equation \textit{per se} are not needed to derive these equations \cite{foot:vlasov}.

\subsection{Longitudinal electrostatic waves}
\label{app:lew}

Finally, let us consider longitudinal electrostatic waves in somewhat more detail. In this case, for the longitudinal polarizability of an individual particle with OC velocity $\vec{V}$, we take $\alpha_s = - e_s^2/[m_s(\omega - \vec{k}\cdot\vec{V})^2]$ \cite{my:mneg}, where $e_s$ and $m_s$ are the particle charge and mass, respectively; in particular, this corresponds to
\begin{gather}\label{eq:lponder}
\Phi_s = \frac{e^2_s |\tilde{\vec{E}}|^2}{4m_s(\omega - \vec{k}\cdot\vec{V})^2}
\end{gather}
(cf. \Refs{ref:cary81, ref:bauer95, my:mneg}). Then, the longitudinal dielectric function,
\begin{gather}\label{eq:ldf}
\epsilon = 1 + 4\pi \sum_s n_s \favr{\alpha_s}_{\avp},
\end{gather}
can be written as (cf. \Ref{ref:bohm49})
\begin{gather}\label{eq:bg}
\epsilon = 1 - \sum_s \omega_{ps}^2 \int^\infty_{-\infty} \frac{f_s(V_x)}{(\omega - kV_x)^2}\,dV_x,
\end{gather}
where $\omega_{ps}^2 = 4\pi n_s e^2_s/m_s$, and $f_s(V_x)$ are the distributions of the particle longitudinal velocities $V_x$, normalized such that $\int^\infty_{-\infty} f_s(V_x)\,dV_x = 1$. By definition, a linear wave has no trapped particles, so $f_s(V_x)$ is zero in the resonance vicinity, and thus the integrand in \Eq{eq:bg} is analytic. Hence, one can take the integral by parts. This yields
\begin{gather}
\epsilon = 1 + \sum_s \frac{\omega_{ps}^2}{k} \int^\infty_{-\infty} \frac{f_s'(V_x)}{\omega - kV_x}\,dV_x,
\end{gather}
in agreement with \Ref{book:landau10}.

From, \Eq{eq:ldre} the dispersion relation now reads as $\epsilon(\omega, \vec{k}) = 0$. In particular, this means $\vec{v}_{g0} = - \epsilon_{\vec{k}}/\epsilon_\omega$ and 
\begin{gather}\label{eq:lewact}
\mc{I} = \frac{\epsilon_\omega}{16\pi}\,|\tilde{\vec{E}}|^2.
\end{gather}
Let us show that this expression is consistent with \Eq{eq:ipt}. First, notice that \cite{my:kchi}
\begin{gather}
P_x = m_sV_x - \pd_{V_x} \Phi_s,
\end{gather}
so $\mc{I} = k^{-1}\sum_s n_s\favr{\pd_{V_x} \Phi_s}_{\avp}$. From \Eq{eq:lponder}, one gets
\begin{gather}
k^{-1} \favr{\pd_{V_x} \Phi_s}_{\avp} = - \favr{\pd_\omega \Phi_s}_{\avp} = - \pd_\omega \favr{\Phi_s}_{\avp},
\end{gather}
and therefore $\mc{I} = - \pd_\omega \sum_s n_s \favr{\Phi_s}_{\avp}$, or
\begin{gather}
\mc{I} = \frac{|\tilde{\vec{E}}|^2}{16\pi}\, \pd_\omega \sum_s 4\pi n_s \favr{\alpha_s}_{\avp},
\end{gather}
where we substituted \Eq{eq:ponder} for $\Phi_s$. Using \Eq{eq:ldf}, one thereby matches \Eq{eq:lewact}, as anticipated.


\begin{thebibliography}{10}

\bibitem{ref:whitham65}
G.~B. Whitham, J. Fluid Mech. {\bf 22}, 273 (1965).

\bibitem{book:whitham}
G.~B. Whitham, {\it Linear and Nonlinear Waves\/} (Wiley, New York, 1974).

\bibitem{ref:lighthill65b}
M.~J. Lighthill, J. Inst. Math. Appl. {\bf 1}, 269 (1965).

\bibitem{ref:bretherton69}
F.~P. Bretherton and C.~J.~R. Garrett, Proc. Roy. Soc. A {\bf 302}, 529 (1968).

\bibitem{ref:dewar70}
R.~L. Dewar, Phys. Fluids {\bf 13}, 2710 (1970).

\bibitem{ref:hayes73}
W.~D. Hayes, Proc. R. Soc. Lond. A. {\bf 332}, 199 (1973).

\bibitem{ref:kravtsov74}
Yu.~A. Kravtsov, L.~A. Ostrovsky, and N.~S. Stepanov, Proc. IEEE {\bf 62}, 1492
  (1974).

\bibitem{ref:dewar77}
R.~L. Dewar, Aust. J. Phys. {\bf 30}, 533 (1977).

\bibitem{ref:hirota10}
M.~Hirota and S.~Tokuda, Phys. Plasmas {\bf 17}, 082109 (2010).

\bibitem{foot:robust}
For example, see \Refs{my:dense, my:mquanta} for discussion on waves in
  laboratory and cosmological plasmas, including those in curved spacetime
  \cite{my:metric, ref:heintzmann83, ref:kulsrud92}.

\bibitem{ref:dewar72}
R.~L. Dewar, Astrophys. J. {\bf 174}, 301 (1972).

\bibitem{ref:dewar72c}
R.~L. Dewar, J. Plasma Phys. {\bf 7}, 267 (1972).

\bibitem{ref:brizard95}
A.~J. Brizard and A.~N. Kaufman, Phys. Rev. Lett. {\bf 74}, 4567 (1995).

\bibitem{my:bgk}
I.~Y. Dodin and N.~J. Fisch, Phys. Rev. Lett. {\bf 107}, 035005 (2011).

\bibitem{ref:bernstein57}
I.~B. Bernstein, J.~M. Greene, and M.~D. Kruskal, Phys. Rev. {\bf 108}, 546
  (1957).

\bibitem{ref:ng06}
C.~S. Ng, A.~Bhattacharjee, and F.~Skiff, Phys. Plasmas {\bf 13}, 055903
  (2006).

\bibitem{ref:schamel00}
H.~Schamel, Phys. Plasmas {\bf 7}, 4831 (2000).

\bibitem{my:bgknum}
P.~F. Schmit, I.~Y. Dodin, and N.~J. Fisch, Phys. Plasmas {\bf 18}, 042103
  (2011).

\bibitem{ref:krasovskii89}
V.~L. Krasovskii, Zh. Eksp. Teor. Fiz. {\bf 95}, 1951 (1989) [Sov. Phys. JETP
  {\bf 68}, 1129 (1989)].

\bibitem{ref:krasovskii95}
V.~L. Krasovskii, Zh. Eksp. Teor. Fiz. {\bf 107}, 741 (1995) [JETP {\bf 80},
  420 (1995)].

\bibitem{ref:benisti07}
D.~B\'enisti and L.~Gremillet, Phys. Plasmas {\bf 14}, 042304 (2007).

\bibitem{ref:matveev09}
A.~I. Matveev, Rus. Phys. J. {\bf 52}, 885 (2009).

\bibitem{ref:bohm49}
D.~Bohm and E.~P. Gross, Phys. Rev. {\bf 75}, 1851 (1949).

\bibitem{ref:kruer69}
W.~L. Kruer, J.~M. Dawson, and R.~N. Sudan, Phys. Rev. Lett. {\bf 23}, 838
  (1969).

\bibitem{ref:goldman71}
M.~V. Goldman and H.~L. Berk, Phys. Fluids {\bf 14}, 801 (1971).

\bibitem{ref:krasovsky94}
V.~L. Krasovsky, Phys. Scripta {\bf 49}, 489 (1994).

\bibitem{ref:krasovsky09}
V.~L. Krasovsky, Plasma Phys. Control. Fusion {\bf 51}, 115011 (2009).

\bibitem{tex:myactii}
I. Y. Dodin and N. J. Fisch, \textit{Adiabatic nonlinear waves with trapped
  particles: II. Wave dispersion} (Paper~II), submitted together with the present paper.

\bibitem{tex:myactiii}
I. Y. Dodin and N. J. Fisch, \textit{Adiabatic nonlinear waves with trapped
  particles: III. Wave dynamics} (Paper~III), submitted together with the present paper.

\bibitem{foot:similar}
Somewhat similar Lagrangians also appeared, \eg in \Refs{ref:dewar72c, ref:dewar77, ref:armstrong75, ref:kulsrud92, ref:brizard00, ref:brizard09}.

\bibitem{foot:modes}
In fact, considered in \Ref{my:bgk} were wave \textit{modes}, so averaging over
  space was also assumed.

\bibitem{book:goldstein}
H.~Goldstein, {\it Classical Mechanics\/} (Addison-Wesley, Reading,~MA, 1950), Sec.~11.5.

\bibitem{book:stix}
T.~H. Stix, {\it Waves in Plasmas\/} (AIP, New York, 1992).

\bibitem{arX:mycoin}
I. Y. Dodin, arXiv:1107.2852v1.

\bibitem{foot:scalar}
We assume the notation $\vec{a} \cdot \vec{b} = \sum_k a_k b_k$.

\bibitem{book:landau1}
L.~D. Landau and E.~M. Lifshitz, {\it Mechanics\/} (Butterworth-Heinemann,
  Oxford, 1976).

\bibitem{foot:avrn}
Since $\mc{H}_i$ depend on $\vec{x}$ only through parameters that vary slowly
  in space, this means that $n_s$ will hence be densities \textit{locally
  averaged} in space. This distinction is important for trapped particles,
  whose OC true densities have a spatial period equal to $\lambda$. Unlike
  those true densities, $n_s$ defined here vary on a much larger scale, namely
  that of the envelope inhomogeneity.

\bibitem{foot:xt}
Explicit dependence on $(t, \vec{x})$ is allowed too, albeit not emphasized
  here for brevity.

\bibitem{ref:seliger68}
R.~L. Seliger and G.~B. Whitham, Proc. Roy. Soc. A {\bf 305}, 1 (1968).

\bibitem{ref:webb05}
G.~M. Webb, G.~P. Zank, E.~Kh. Kaghashvili, and R.~E. Ratkiewicz, J. Plasma
  Phys. {\bf 71}, 785 (2005).

\bibitem{ref:webb07}
G.~M. Webb, E.~Kh. Kaghashvili, and G.~P. Zank, J. Plasma Phys. {\bf 73}, 15
  (2007).

\bibitem{foot:high}
Of course, one could also make the formulation more precise, by allowing $\mcc{L}$ to depend on derivatives of $a$ too; cf. Ref.~\cite[Sec.~15.5]{book:whitham}.

\bibitem{my:mneg}
I.~Y. Dodin and N.~J. Fisch, Phys. Rev. E {\bf 77}, 036402 (2008).

\bibitem{foot:notenough}
However, when calculating the nonlinear frequency shift, the nonlinear $\mc{E}$
  must be retained for both passing and trapped particles, even at small $a$
  \cite{my:bgk, tex:myactii}.

\bibitem{ref:benisti10}
D.~B\'enisti, O.~Morice, L.~Gremillet, E.~Siminos, and D.~J. Strozzi, Phys.
  Plasmas {\bf 17}, 082301 (2010).

\bibitem{foot:impressed}
Another signal velocity is $u$, which is the propagation velocity for modulations impressed on $\sigma_s$. 

\bibitem{foot:landau}
In principle, collisionless dissipation can be included in a Hamiltonian
  formulation \cite{ref:escande96}, as well as any other dissipative effects
  \cite{ref:jimenez76}.

\bibitem{book:arnold06}
V.~A. Arnold, V.~V. Kozlov, and A.~I. Neishtadt, {\it Mathematical Aspects of
  Classical and Celestial Mechanics\/} (Springer, New York, 2006), Sec.~3.2.

\bibitem{my:kchi}
I.~Y. Dodin and N.~J. Fisch, Phys. Lett. A {\bf 374}, 3472 (2010).

\bibitem{my:nlinphi}
I.~Y. Dodin and N.~J. Fisch, Phys. Rev. E {\bf 79}, 026407 (2009).

\bibitem{ref:larsson86}
J.~Larsson, J. Math. Phys. {\bf 27}, 495 (1986).

\bibitem{ref:pfirsch04}
D.~Pfirsch and D.~Correa-Restrepo, J. Plasma Phys. {\bf 70}, 719 (2004).

\bibitem{foot:vlasov}
The Vlasov equation is not needed because $\oper{\epsilon}$ is expressed in
  terms of $\oper{\alpha}_s$, which are found from the single-particle motion
  equations; \eg see \App{app:lew}.

\bibitem{ref:cary81}
J.~R. Cary and A.~N. Kaufman, Phys. Fluids {\bf 24}, 1238 (1981).

\bibitem{ref:bauer95}
D.~Bauer, P.~Mulser, and W.~H. Steeb, Phys. Rev. Lett. {\bf 75}, 4622 (1995).

\bibitem{book:landau10}
E.~M. Lifshitz and L.~P. Pitaevskii, {\it Physical Kinetics\/} (Pergamon Press,
  New York, 1981), Sec.~29.

\bibitem{my:dense}
I.~Y. Dodin, V.~I. Geyko, and N.~J. Fisch, Phys. Plasmas {\bf 16}, 112101
  (2009).

\bibitem{my:mquanta}
I.~Y. Dodin and N.~J. Fisch, Phys. Rev. D {\bf 82}, 044044 (2010).


\bibitem{my:metric}
I.~Y. Dodin and N.~J. Fisch, Phys. Plasmas {\bf 17}, 112118 (2010).

\bibitem{ref:heintzmann83}
H.~Heintzmann and M.~Novello, Phys. Rev. A {\bf 27}, 2671 (1983).

\bibitem{ref:kulsrud92}
R.~Kulsrud and A.~Loeb, Phys. Rev. D {\bf 45}, 525 (1992). 

\bibitem{ref:armstrong75}
J.~A. Armstrong, Phys. Rev. A {\bf 11}, 963 (1975).

\bibitem{ref:brizard00}
A.~J. Brizard, Phys. Rev. Lett. {\bf 84}, 5768 (2000).

\bibitem{ref:brizard09}
A.~J. Brizard, J. Phys. Conf. Ser. {\bf 169}, 012003 (2009).

\bibitem{ref:escande96}
D.~F. Escande, S.~Zekri, and Y.~Elskens, Phys. Plasmas {\bf 3}, 3534 (1996).

\bibitem{ref:jimenez76}
J.~Jimenez and G.~B. Whitham, Proc. R. Soc. A {\bf 349}, 277 (1976).

\end{thebibliography}
\end{document}